\documentclass{PoS}

\newcommand{\bm}[1]{\mbox{\boldmath$#1$}}

\title{Spectral Properties of Quarks in the Quark-Gluon Plasma}

\ShortTitle{Spectral Properties of Quarks in the Quark-Gluon Plasma}

\author{Frithjof Karsch\\
        Brookhaven National Laboratory, Bldg.510A, Upton, 11973, NY, USA\\
        E-mail: \email{karsch@quark.phy.bnl.gov}}

\author{\speaker{Masakiyo Kitazawa}\\ 
        Department of Physics, Osaka University, Toyonaka, Osaka, 560-0043, Japan\\
        E-mail: \email{kitazawa@phys.sci.osaka-u.ac.jp}}

\abstract{
We analyze the spectral properties of the quark propagator 
above the critical temperature for the deconfinement phase 
transition in quenched lattice QCD using clover improved 
Wilson fermions. The bare quark mass dependence of the quark 
spectral function is analyzed by varying the hopping parameter 
$\kappa$ in Landau gauge. We assume a two-pole structure 
for the quark spectral function, which is numerically found 
to work quite well for any value of $\kappa$. It is shown 
that in the chiral limit the quark spectral function has 
two collective modes that correspond to the normal and 
plasmino excitations, while it is dominated by a single-pole 
structure when the bare quark mass becomes large.
}

\FullConference{The XXV International Symposium on Lattice Field Theory\\
		 July 30-4 August 2007\\
		 Regensburg, Germany}

\begin{document}

\section{Introduction}

Hadronic matter undergoes a phase transition to 
a deconfined phase at nonzero temperature.
The properties of the phase above the critical temperature, 
$T_c$, acquires much interest
both from the experimental and theoretical point of view.
In order to understand the structure of the matter
in this region, it is desirable to identify the basic 
degrees of freedom of the system and their quasi-particle 
properties. 
In the present study, we analyze dynamical properties of 
quarks above $T_c$ in quenched lattice QCD in Landau gauge
\cite{Karsch:2007wc}.

At asymptotically high temperatures 
one can calculate the quark propagator using 
perturbative techniques. 
It is known that the collective excitations of quarks  
in this limit develop a mass gap (thermal mass) 
that is proportional to $gT$ \cite{plasmino,LeBellac}. 
Here $g$ and $T$ denote the gauge coupling and temperature, respectively.
Moreover, in this limit the number of poles in the quark propagator
is doubled. 
In addition to the normal modes,
which reduce to poles in the free particle propagator, 
plasmino modes appear.

In order to understand the origin of the plasmino mode in the quark
propagator in the high temperature limit,
it is instructive to consider the quark propagator 
at intermediate temperature \cite{BBS92,KKN06}.
In \cite{BBS92} the temperature dependence of the 
spectral function for fermions with scalar mass $m$ 
has been considered in QED and a Yukawa model.
In these models the spectral function at zero temperature 
has two poles at energies $\omega=\pm m$,
while in the high temperature limit, $T/m\to\infty$,
it approaches the propagator having four poles.
The one-loop calculation performed in \cite{BBS92} clearly showed
that the two limiting forms of the spectral function are 
connected continuously; 
in addition to the normal quasi-particle peak 
a peak corresponding to the plasmino gradually appears
in the spectral function  
and becomes larger with increasing temperature, 
\cite{BBS92}.
In the present study, we analyze the quark propagator 
in quenched lattice QCD
at two values of the temperature, $T=1.5T_c$ and $3T_c$, 
as a function of the bare quark mass.
To simplify the present analysis, all our calculations have been
performed for zero momentum.

\section{Quark Spectral Function}

The dynamical properties of quarks at zero momentum are 
encoded in the quark spectral function $\rho( \omega )$
which is related to the Euclidean correlation function
\begin{eqnarray}
S( \tau ) 
= \frac1V \int d^3 x d^3 y 
\langle \psi( \tau,\bm{x} ) \bar\psi( 0,\bm{y} ) \rangle,
\label{eq:S}
\end{eqnarray}
through an integral equation
\begin{eqnarray}
S( \tau ) = \int_{-\infty}^{\infty}
d\omega \frac{ e^{ (1/2-\tau T)\; \omega/T } }
{ e^{\omega/2T} + e^{-\omega/2T} } \rho( \omega ),
\label{eq:SKrho}
\end{eqnarray}
with the quark field $\psi$, the spatial volume $V$, and
the imaginary time $\tau$ which is restricted to $0\le \tau<1/T$.
The Dirac structure of $\rho( \omega )$ is decomposed as
\begin{eqnarray}
\rho( \omega )
&=& \rho_{\rm 0} ( \omega ) \gamma^0 + \rho_{\rm s} ( \omega )
\nonumber \\
&=&\rho_+ ( \omega ) \Lambda_+ \gamma^0 
+ \rho_- ( \omega ) \Lambda_- \gamma^0,
\label{eq:rho}
\end{eqnarray}
with projection operators $ \Lambda_\pm = ( 1 \pm \gamma^0 )/2 $.
The charge conjugation symmetry leads to 
$\rho_{\rm 0} ( \omega ) =  \rho_{\rm 0} ( -\omega )$,
$\rho_{\rm s} ( \omega ) =  -\rho_{\rm s} ( -\omega )$, and
$\rho_+ (\omega) = \rho_- (-\omega)
 = \rho_{\rm 0} (\omega) + \rho_{\rm s} (\omega)$
\cite{BBS92}.
In the following analysis, we concentrate on a determination of 
$\rho_\pm(\omega)$ instead of $\rho_{\rm 0,s}(\omega)$,
because excitation properties of quarks are more apparent
in these channels. 
The spectral functions $\rho_\pm (\omega)$
are neither even nor odd functions. In the chiral limit, however,
$\rho_{\rm s}$ vanishes and $\rho_\pm(\omega)$ become even functions.
In analogy to Eq.~(\ref{eq:rho}) we introduce the decomposition of the
correlation function $S( \tau )$ as $S (\tau)
=S_+ (\tau) \Lambda_+ \gamma^0 + S_- (\tau) \Lambda_- \gamma^0$, where
$S_\pm$ are related through $S_+(\tau) = S_-(\beta-\tau)$.

\begin{figure}
\begin{center}
\includegraphics[width=.55\textwidth]{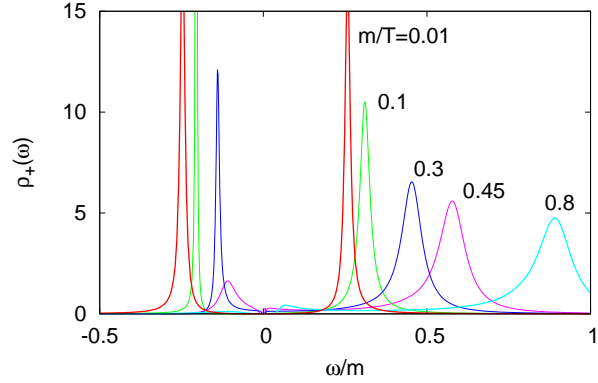}
\caption{
The spectral function $\rho_+(\omega)$
in the Yukawa model with massive fermion and massless boson
for various values of $m/T$ \cite{BBS92}.
}
\label{fig:yukawa}
\end{center}
\end{figure}

For free quarks with scalar mass $m$ the spectral functions,
$\rho_\pm (\omega) = \pi \delta (\omega \mp m)$, 
have quark and anti-quark poles 
at $\omega= \pm m$, respectively.  
In the high temperature limit, additional poles,
corresponding to the plasmino, appear at negative energy for 
$\rho_+ (\omega)$ and positive energy for $\rho_- (\omega)$ 
\cite{BBS92,KKN06}.
In Fig.~\ref{fig:yukawa}, we show $\rho_+ (\omega)$
in the Yukawa model with a massless boson at finite $T$
for various values of $m/T$ \cite{BBS92,KKprep}.
One sees that the shape of $\rho_+ (\omega)$ changes
continuously between these two limits as $m/T$ varies.

To extract the spectral function $\rho_+(\omega)$ from $S(\tau)$
using Eq.~(\ref{eq:SKrho}), we assume that $\rho_+(\omega)$ 
can be described by a two-pole ansatz,
\begin{eqnarray}
\rho_+(\omega)
= Z_1 \delta( \omega - E_1 ) + Z_2 \delta( \omega + E_2 ),
\label{eq:2pole}
\end{eqnarray}
where the residues $Z_{1,2}$ and energies $E_{1,2}>0$ have to be
determined from a fit to $S_+(\tau)$. 
The poles at $\omega=E_1, -E_2$ correspond to
the normal and plasmino modes, respectively
\cite{BBS92}.

\begin{table}
\begin{center}
\begin{tabular}{ccrcccc}
\hline
\hline
$T/T_c$ & $N_\tau$ & $N_\sigma$ & $\beta$ & $c_{\rm SW}$ & $\kappa_c$ & $a$[fm] \\
\hline
$3$     &    $16$ & $64,48$     & $7.457$ & $1.3389$     & $0.13390$  & $0.015$ \\
        &    $12$ & $48$        & $7.192$ & $1.3550$     & $0.13437$  & $0.021$ \\
$1.5$   &    $16$ & $64,48$     & $6.872$ & $1.4125$     & $0.13495$  & $0.031$ \\
        &    $12$ & $48$        & $6.640$ & $1.4579$     & $0.13536$  & $0.041$ \\
\hline
\hline
\end{tabular}
\end{center}
\caption{
Simulation parameters \cite{params}.
}
\label{table}
\end{table}

The correlation function $S(\tau)$ has been calculated at two
values of the temperature, $T =1.5T_c$ and $3T_c$, in 
quenched QCD using non-perturbatively improved clover Wilson 
fermions \cite{Sheikholeslami:1985ij,Luscher:1996jn}.
To control the dependence of our results on the finite lattice 
volume, $N_\sigma^3 \times N_\tau$, and lattice spacing, $a$,
we analyze the quark propagator on lattices of three different 
sizes. 
The gauge field ensembles used for this analysis have been
generated and used previously by the Bielefeld group to study
screening masses and spectral functions  \cite{params}.
The different simulation
parameters are summarized in Table~\ref{table} \cite{params}.
For each lattice size, $51$ configurations have been analyzed. 
Quark propagators have been calculated after fixing each 
gauge field configuration to Landau gauge. For this we used
a conventional minimization algorithm with a stopping 
criterion, $(1/3){\rm tr}|\partial_\mu A^\mu|^2 <10^{-11}$.
In the Wilson fermion formulation the 
bare mass, $m_0$, is related to the  hopping parameter $\kappa$,
through the standard relation
$
m_0 = \left( 1/\kappa - 1/{\kappa_c} \right)/(2a) ,
$
where $\kappa_c$ denotes the critical hopping parameter corresponding
to the chiral limit, or vanishing quark mass.

\section{Numerical Results}

\begin{figure}[tbp]
\begin{center}
\includegraphics[width=.55\textwidth]{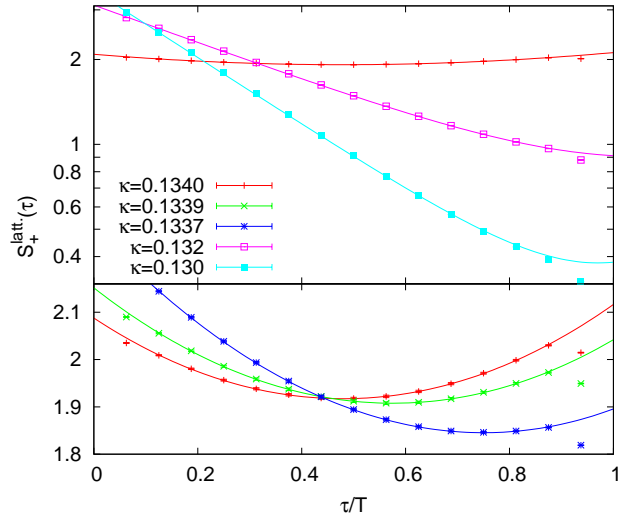}
\caption{
The lattice correlation function $S_+^{\rm latt.}(\tau)$
at $T=3T_c$ for the lattice of size $64^3\times16$
with various values of $\kappa$, and the 
fitting result with the ansatz 
Eq.~(2.4). 
}
\label{fig:S}
\end{center}
\end{figure}

In Fig.~\ref{fig:S}, we show the numerical results
for $S_+^{\rm latt.}(\tau)$ for several values of $\kappa$
calculated on a lattice of 
size $64^3\times16$ at $T=3T_c$.
One sees that the shape of $S_+^{\rm latt.}(\tau)$
approaches that of a single exponential function for smaller $\kappa$,
while it becomes symmetric as $\kappa$ approaches $\kappa_c$.
In the vicinity of the wall source, {\it i.e.} at small and large
$\tau$, we see deviations from this generic picture which can
be attributed to distortion effects arising from the presence of the
source.
We thus exclude points with $\tau < \tau_{min}$ and
$N_\tau -\tau < \tau_{min}$ from our fits to the ansatz
given in Eq.~(\ref{eq:2pole}).
The resulting correlation functions obtained from correlated fits
with $\tau_{min}=3$ are shown in Fig.~\ref{fig:S}.
One sees that $S_+^{\rm latt.}$ is well reproduced by
our fitting ansatz\footnote{We also checked that fits 
based only on a single pole ansatz lead to 
unacceptable large  $\chi^2/{\rm dof}$.};
the $\chi^2/{\rm dof}$
of our fits is between $2$ and $3$ at 
$1.335\lesssim\kappa\lesssim1.34$, 
while it gradually increases as $\kappa$ becomes smaller 
than $\kappa=1.335$.
A similar behavior is also observed for our other lattice sizes \cite{KKprep}.

\begin{figure}[tbp]
\begin{center}
\includegraphics[width=.55\textwidth]{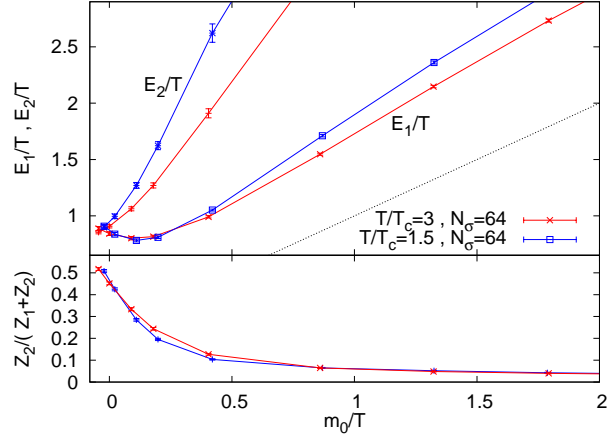}
\caption{
The bare quark mass dependence of fitting parameters 
$E_{1,2}$ and $Z_2 / ( Z_1+Z_2 )$ 
at $T=1.5T_c$ and $3T_c$ for lattice $64^3\times16$.
}
\label{fig:ZE}
\end{center}
\end{figure}

In Fig.~\ref{fig:ZE}, we show the dependence of 
$E_{1,2}$ and $Z_2 / ( Z_1+Z_2 )$ on the bare quark
mass $m_0$ for $T=1.5T_c$ and $3T_c$. 
The results have been obtained
from two-pole fits on lattices of size $64^3\times16$.
Errorbars have been estimated from a  Jackknife analysis.
The dotted line in this figure denotes the pole mass determined
from the bare lattice mass.
The figure shows that
the ratio $Z_2 / (Z_1+Z_2)$ becomes larger with
decreasing $m_0$ and eventually reaches $0.5$.
The hopping parameters satisfying $Z_1=Z_2$ are
$\kappa'_c = 0.133974(10)$ for $T=3T_c$ and
$\kappa'_c = 0.134991(9)$ for $T=1.5T_c$, 
which are consistent with the values for
$\kappa_c$ given in Table~\ref{table}.
The latter had been obtained in \cite{params} from a fit 
to critical hopping parameters determined in \cite{Luscher:1996jn} 
from the vanishing of the isovector axial current.
The numerical results obtained on $64^3\times16$ lattices 
show that $E_1$ and $E_2$ are equal 
within statistical errors at $\kappa=\kappa'_c$.
The spectral function $\rho_+(\omega)$ thus becomes an 
even function at this point; the quark propagator becomes
chirally symmetric despite the presence of a thermal mass,
$m_T \equiv E_1=E_2$.
From Fig.~\ref{fig:ZE}, one also finds that the ratio $m_T/T$
is insensitive to $T$ in the temperature range analyzed in this work,
while it is slightly larger for lower $T$.

As $m_0$ becomes larger, $Z_2/(Z_1+Z_2)$ decreases and 
$\rho_+(\omega)$ is eventually dominated by a single-pole.
One sees that $E_1$ has a minimum at $m_0>0$,
while $E_2$ is an increasing function of $m_0$.
In the one-loop approximation, the peaks in $\rho_+(\omega)$
corresponding to $E_1$ ($E_2$) are monotonically
increasing (decreasing) function of $m_0/T$ \cite{BBS92,KKprep}.
The quark mass dependence of poles found here thus is
qualitatively different from the perturbative result.
We find, however, that slope of $E_2$ as function of $m_0/T$
decreases with increasing $T$. This may suggest that the
perturbative behavior could eventually be recovered at
much larger temperatures.

\begin{figure}[tbp]
\begin{center}
\includegraphics[width=.5\textwidth]{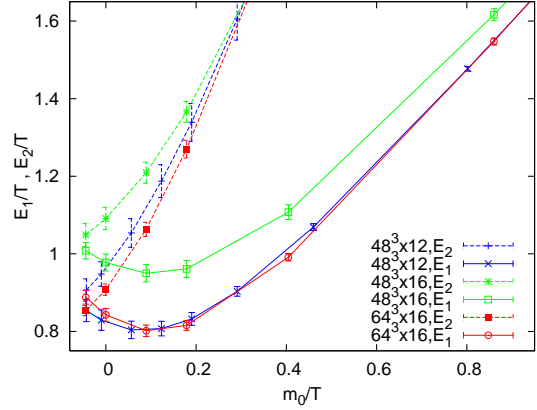}
\caption{
The bare quark mass dependence of parameters $E_1$, $E_2$ 
at $T=3T_c$ for lattices of size
$64^3\times16$, $48^3\times16$ and $48^3\times12$.
}
\label{fig:compare}
\end{center}
\end{figure}

In order to check the dependence of our results on the lattice 
spacing and finite volume, we analyzed the quark propagator 
at $T=3T_c$ for three different lattice sizes. Results for
$E_1$ and $E_2$ are shown in Fig.~\ref{fig:compare}. 
Comparing the results obtained on lattices with different 
lattice cut-off, $a$,
but same physical volume, {\it i.e.} $64^3\times16$ and 
$48^3\times12$, one sees that any possible cut-off dependence 
is statistically not significant in our analysis. On the other
hand we find a clear dependence of the quark 
energy levels on the spatial volume; when comparing lattices
with aspect ratio $N_\sigma/N_\tau =3$ and $4$ we find that the
energy levels, $E_{1,2}$, drop significantly. 
A similar behavior is observed also at $T=1.5T_c$.

\begin{figure}[tbp]
\begin{center}
\includegraphics[width=.5\textwidth]{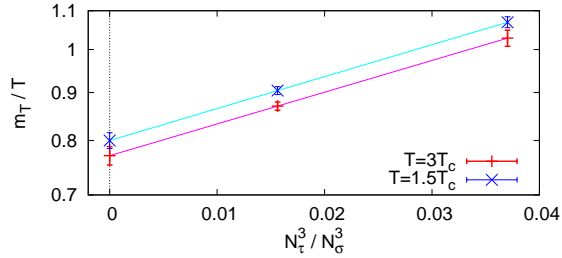}
\caption{
The bare quark mass dependence of fitting parameters 
$E_{1,2}$ and $Z_2 / ( Z_1+Z_2 )$ 
at $T=1.5T_c$ and $3T_c$ for lattice $64^3\times16$.
}
\label{fig:extrapl}
\end{center}
\end{figure}

The presence of a strong volume dependence of the quark propagator
is not unexpected. In fact, the thermal quark mass arises as 
collective effect of low momentum gluons; gluons at the soft 
scale $p\lesssim gT$ play a crucial role to give rise to the thermal 
mass at high temperatures \cite{LeBellac}. However, on lattices 
with given aspect ratio $N_\sigma/N_\tau$ low momentum gluons are 
cut-off. The lowest non-vanishing gluon momentum is,
$p_{min}/T = 2\pi (N_\tau / N_\sigma)$, which still is larger than
unity on lattices with aspect ratio $N_\sigma/N_\tau=4$. 
The situation may, nonetheless, be somewhat better in the
temperature range explored here as the temperature 
dependent coupling $g(T)$ is larger than unity.
An analysis of quark spectral functions on lattices
with even larger spatial volume is needed in the future to
properly control effects of small momenta. 
We attempted to estimate the thermal mass in the $V\to\infty$ limit
by extrapolating the results obtained for two different volumina.
Defining $m_T \equiv ( Z_1 E_1 + Z_2 E_2 ) / ( Z_1+Z_2) 
|_{\kappa = \kappa_c}$ and assuming
the volume dependence of $m_T$ as 
$m_T(N_\tau/N_\sigma)= m_T(0) \exp(N_\tau^3/N_\sigma^3)$,
we obtain $m_T(0)/T = 0.771(18)$ for $T=3T_c$
and $m_T(0)/T = 0.800(15)$ for $T=1.5T_c$.
This suggests that finite volume effects may still be 
of the order of 15\% in our current analysis of $m_T/T$  
(see Fig.~\ref{fig:extrapl}).
Despite these problems, our result clearly shows that
light quarks near but above $T_c$ have a mass gap that is 
of collective nature similar to that in the perturbative regime.

\section{Summary}

In this study,
we analyzed the quark spectral function at zero momentum 
for $T=1.5T_c$ and $3T_c$ as functions of bare quark mass $m_0$ 
in quenched lattice QCD with Landau gauge fixing.
We found that the two-pole approximation for $\rho_+(\omega)$
well reproduces the behavior of the lattice correlation function.
It is argued that the chiral symmetry of the quark propagator is
restored at the critical value of $\kappa$ and
the shape of the spectral function at this point takes
a similar form as in the high temperature limit
having normal and plasmino modes with thermal mass $m_T$.
Meanwhile, $\rho_+(\omega)$ approaches a single-pole structure
as $m_0$ is increased, as one can naturally deduce intuitively.
The non-perturbative nature of thermal gauge fields is reflected 
in the behavior of poles as functions of $m_0$, which is 
qualitatively different from the perturbative result 
\cite{BBS92}.
We also note that the ratio $m_T/T$ decreases slightly
with increasing $T$, which is expected to happen at
high temperature where $m_T/T$ should be proportional to 
a running coupling $g(T)$.
Although results on the quark propagator are gauge dependent,
we expect that our results for its poles suffer less from gauge 
dependence,
because the success of the pole approximation for $\rho_+(\omega)$
indicates that the quark propagator 
has dynamical poles near the real axis, which are 
gauge independent quantities. 

In the present study, we analyzed the quark spectral function
in the quenched approximation.
Although this approximation includes the leading contribution
in the high temperature limit \cite{LeBellac} and thus is valid 
at sufficiently high $T$, 
the validity of this approximation near $T_c$ is nontrivial.
For example, screening of gluons due to the 
polarization of the vacuum with virtual quark antiquark pairs
is neglected in this approximation.
The coupling to possible mesonic excitations,
which may cause interesting effects in the spectral properties
of the quark \cite{KKN06}, are not incorporated, either.
The comparison of the quark propagator between quenched and
full lattice simulations would tell us 
the strength of these effects near $T_c$.

\end{document}